\documentstyle[12pt]{article}
\begin{document}
\setlength{\baselineskip}{23pt}
\newcommand{\vs}{\vspace{0.2cm}}
\newcommand{\vsa}{\vspace{0.4cm}}
\newcommand{\vsb}{\vspace{0.8cm}}
\newcommand{\invexp}{\rule{0pt}{10pt}}
\newcommand{\invexpa}{\rule{0pt}{20pt}}

\begin{titlepage}
\begin{flushright} UCLA/92/TEP/22 \\ May 1992 \end{flushright}
\vsb
\begin{center}
{\large\bf Bounds from Stability and Symmetry Breaking on\vs\\
Parameters in the Two-Higgs-Doublet Potential\vsa\vsb\\}
Boris Kastening\footnote{email: boris@ruunts.fys.ruu.nl}\vsb\\
{\it Department of Physics\\
University of California, Los Angeles\\
Los Angeles, California 90024}\vsb\vsb\\
{\bf Abstract}\\
\end{center}
The most general ${\rm SU}(2)\times{\rm U}(1)_Y$-symmetric quartic potential
with two Higgs doublets, subject to an only softly broken discrete symmetry
$(\phi_1,\phi_2)\rightarrow(-\phi_1,\phi_2)$, is considered. At tree-level,
analytic bounds on the parameters are derived that ensure a stable vacuum,
breaking ${\rm SU}(2)\times{\rm U}(1)_Y$ down to ${\rm U}(1)_{\rm em}$.
\end{titlepage}

\begin{sloppypar}
In the minimal standard model, spontaneous symmetry breaking of the
\mbox{${\rm SU}(2)\times{\rm U}(1)_Y$} gauge symmetry down to
${\rm U}(1)_{\rm em}$ is caused by the potential of a single
Higgs-${\rm SU}(2)$ doublet. However, since the exact symmetry
breaking mechanism is not known, it is natural to consider extensions
of the Higgs sector. The simplest such extension is to have another
${\rm SU}(2)$-doublet scalar field. In fact, there are many reasons
to discuss the two-doublet model. For associated particle phenomenology,
see e.g.\ \cite{GuHa..,Sh} and references therein. More recently, the model
has also received attention in cosmological contexts, mainly in connection
with baryogenesis in the early universe \cite{BoKu..}.
\end{sloppypar}

Assume we want to prescribe the minimum of the two-doublet potential as
\begin{equation}
\label{dbmin}
<\Phi_1>=\frac{1}{\sqrt{2}}\left(\begin{array}{c}0\\
v_1\end{array}\right),\;\;\;\;\;\;
<\Phi_2>=\frac{e^{i\rho}}{\sqrt{2}}\left(\begin{array}{c}0\\
v_2\end{array}\right),
\end{equation}
which, up to gauge transformations, is the most general vacuum that breaks
${\rm SU}(2)\times{\rm U}(1)_Y$ down to ${\rm U}(1)_{\rm em}$.
Then, up to an overall constant, the most general
${\rm SU}(2)\times{\rm U}(1)_Y$-symmetric quartic potential involving two
Higgs-${\rm SU}(2)$ doublets that is subject to an only softly (i.e.\ by
dimension-two terms) broken discrete symmetry
$(\phi_1,\phi_2)\rightarrow(-\phi_1,\phi_2)$ [equivalent to only soft
breaking of $(\phi_1,\phi_2)\rightarrow(\phi_1,-\phi_2)$] and that makes
(\ref{dbmin}) stationary, can be written as\footnote{I am grateful to
Howard Haber for pointing out to me the possibility of having the
$\lambda_7$-term.}
\begin{eqnarray}
\label{dbpotential}
V(\Phi_1,\Phi_2)=\!\!\!\!
&&\!\!\!\!\begin{array}{l}\lambda_1(\Phi_1^\dagger\Phi_1-\frac{v_1^2}{2})^2
+\lambda_2(\Phi_2^\dagger\Phi_2-\frac{v_2^2}{2})^2\end{array}\nonumber\\
&+&\!\!\!\!\begin{array}{l}\lambda_3[(\Phi_1^\dagger\Phi_1-\frac{v_1^2}{2})
+(\Phi_2^\dagger\Phi_2-\frac{v_2^2}{2})]^2\end{array}\nonumber\\
&+&\!\!\!\!\begin{array}{l}\lambda_4
[(\Phi_1^\dagger\Phi_1)(\Phi_2^\dagger\Phi_2)
-(\Phi_1^\dagger\Phi_2)(\Phi_2^\dagger\Phi_1)]\end{array}\nonumber\\
&+&\!\!\!\!\begin{array}{l}\lambda_5[{\rm Re}(\Phi_1^\dagger\Phi_2)
-\frac{v_1v_2}{2}\cos\rho]^2+\lambda_6[{\rm Im}(\Phi_1^\dagger\Phi_2)
-\frac{v_1v_2}{2}\sin\rho]^2\end{array}\nonumber\\
&+&\!\!\!\!\begin{array}{l}\lambda_7[{\rm Re}(\Phi_1^\dagger\Phi_2)
-\frac{v_1v_2}{2}\cos\rho]
[{\rm Im}(\Phi_1^\dagger\Phi_2)-\frac{v_1v_2}{2}\sin\rho].\end{array}
\end{eqnarray}
The discrete symmetry imposed on the quartic terms is intended to suppress
neutral flavor-changing currents.

At this point we are free to make a phase redefinition of the $\Phi_i$.
Choosing
$(\Phi_1,\Phi_2)\rightarrow(\Phi_1e^{i\varphi},\Phi_2)$ with $\varphi$
satisfying
\begin{equation}
\label{db567phi}
(\lambda_6-\lambda_5)\sin(2\varphi)=\lambda_7\cos(2\varphi),
\end{equation}
we can write
the potential again in the form (\ref{dbpotential}), with the replacements
\begin{eqnarray}
\lambda_7&\rightarrow&0,\\
\label{db56phi}
\lambda_{5,6}&\rightarrow&\frac{\lambda_5+\lambda_6}{2}
\pm\frac{\lambda_5-\lambda_6}{2\cos2\varphi}
=\frac{\lambda_5+\lambda_6}{2}\mp\frac{\lambda_7}{2\sin2\varphi},\\
\rho&\rightarrow&\rho+\varphi.
\end{eqnarray}
Therefore we will assume $\lambda_7=0$ for the remainder of this paper,
which leaves us with the potential given in \cite{Ge,GuHa..}. Since we
want (\ref{dbmin}) to be the absolute minimum and because clearly $V=0$ for
$\Phi_1$ and $\Phi_2$ given by (\ref{dbmin}), our strategy will be to ask,
for what values of $\lambda_{1,\ldots,6}$ and $\rho$ the potential is
nonnegative for all $\Phi_1$ and $\Phi_2$. After deriving necessary
conditions for $V\geq0$, we will show that they are also sufficient. We will
assume throughout our calculation that $v_1,v_2\neq0$. If one of them
vanishes, the analysis is much easier and leads to the same results.

Take
\begin{equation}
\label{dbphil4}
\Phi_1
=\frac{1}{\sqrt{2}}\left(\begin{array}{c}-cv_2\\v_1\end{array}\right),
\;\;\;\;\;\;
\Phi_2
=\frac{e^{i\rho}}{\sqrt{2}}\left(\begin{array}{c}cv_1\\v_2\end{array}\right),
\end{equation}
with $c$ real. Then
\begin{equation}
V=\frac{1}{4}[\lambda_1v_2^4+\lambda_2v_1^4+\lambda_3(v_1^2+v_2^2)^2
+(\lambda_5\cos^2\!\rho+\lambda_6\sin^2\!\rho)v_1^2v_2^2]c^4
+\frac{1}{4}\lambda_4(v_1^2+v_2^2)^2c^2,
\end{equation}
and we need
\begin{equation}
\label{dbl4nec}
\lambda_4\geq0,
\end{equation}
since otherwise $V<0$ for sufficiently small but nonzero $c^2$.

Take
\begin{equation}
\label{dbphil5a}
\Phi_1
=\frac{1}{\sqrt{2}}\left(\begin{array}{c}0\\v_1\end{array}\right),
\;\;\;\;\;\;
\Phi_2
=\frac{e^{-i\rho}}{\sqrt{2}}\left(\begin{array}{c}0\\-v_2\end{array}\right)
\end{equation}
to get
\begin{equation}
V=\lambda_5v_1^2v_2^2\cos^2\!\rho.
\end{equation}
Therefore we need
\begin{equation}
\label{dbl5nec}
\lambda_5\geq0
\end{equation}
if $\cos\rho\neq0$. If $\cos\rho=0$ take
\begin{equation}
\label{dbphil5b6b}
\Phi_1
=\frac{1}{\sqrt{2}}\left(\begin{array}{c}0\\v_1\end{array}\right),
\;\;\;\;\;\;
\Phi_2
=\frac{e^{i\epsilon}}{\sqrt{2}}\left(\begin{array}{c}0\\v_2\end{array}\right).
\end{equation}
Defining $\delta\equiv|\sin\epsilon-\sin\rho|$ and using $\sin\rho=\pm1$,
we get
\begin{equation}
V=[2\lambda_5\delta+(\lambda_6-\lambda_5)\delta^2]\frac{v_1^2v_2^2}{4}.
\end{equation}
Again we need (\ref{dbl5nec}), since otherwise $V<0$ for sufficiently small
but nonzero $\delta$. Thus, for any $\rho$, (\ref{dbl5nec}) is necessary.

Take
\begin{equation}
\label{dbphil6a}
\Phi_1
=\frac{1}{\sqrt{2}}\left(\begin{array}{c}0\\v_1\end{array}\right),
\;\;\;\;\;\;
\Phi_2
=\frac{e^{-i\rho}}{\sqrt{2}}\left(\begin{array}{c}0\\v_2\end{array}\right).
\end{equation}
to get
\begin{equation}
V=\lambda_6v_1^2v_2^2\sin^2\!\rho.
\end{equation}
Therefore we need
\begin{equation}
\label{dbl6nec}
\lambda_6\geq0
\end{equation}
if $\sin\rho\neq0$. If $\sin\rho=0$, take again $\phi_1$ and $\phi_2$
from (\ref{dbphil5b6b}). Defining $\delta\equiv|\cos\epsilon-\cos\rho|$
and using $\cos\rho=\pm1$, we get
\begin{equation}
V=[2\lambda_6\delta+(\lambda_5-\lambda_6)\delta^2]\frac{v_1^2v_2^2}{4}.
\end{equation}
Again we need (\ref{dbl6nec}), since otherwise $V<0$ for sufficiently small
but nonzero $\delta$. Thus, for any $\rho$, (\ref{dbl6nec}) is necessary.

Set alternatively $\Phi_2=0$ and $\Phi_1=0$. Vacuum stability,
i.e.\ $V\geq0$ for large fields, requires then
\begin{equation}
\label{dbl123nec}
\lambda_1+\lambda_3\geq0,\;\;\;\;\;\;\;\;\lambda_2+\lambda_3\geq0.
\end{equation}

If both $\lambda_1+\lambda_3>0$ and $\lambda_2+\lambda_3>0$, take
\begin{equation}
\label{dbphil123a}
\Phi_1
=\frac{1}{\sqrt{2}}
\left(\frac{\lambda_2+\lambda_3}{\lambda_1+\lambda_3}\right)^\frac{1}{4}
\left(\begin{array}{c}0\\v_2\end{array}\right),
\;\;\;\;\;\;
\Phi_2
=\frac{e^{i\rho}}{\sqrt{2}}
\left(\frac{\lambda_1+\lambda_3}{\lambda_2+\lambda_3}\right)^\frac{1}{4}
\left(\begin{array}{c}0\\v_1\end{array}\right)
\end{equation}
and get
\begin{equation}
V=\frac{1}{2}\left(
1-\frac{\lambda_3}{\sqrt{(\lambda_1+\lambda_3)(\lambda_2+\lambda_3)}}
\right)\left(\sqrt{\lambda_2+\lambda_3}\,v_2^2
-\sqrt{\lambda_1+\lambda_3}\,v_1^2\right)^2.
\end{equation}
Therefore, if
$\sqrt{\lambda_2+\lambda_3}\,v_2^2\neq\sqrt{\lambda_1+\lambda_3}\,v_1^2$,
we need
\begin{equation}
\label{dbl3nec1}
\lambda_3\leq\sqrt{(\lambda_1+\lambda_3)(\lambda_2+\lambda_3)}.
\end{equation}
If $\sqrt{\lambda_2+\lambda_3}\,v_2^2=\sqrt{\lambda_1+\lambda_3}\,v_1^2$,
choose
\begin{equation}
\label{dbphil123b}
\Phi_1
=\frac{c}{\sqrt{2}}
\left(\frac{\lambda_2+\lambda_3}{\lambda_1+\lambda_3}\right)^\frac{1}{4}
\left(\begin{array}{c}0\\v_2\end{array}\right),
\;\;\;\;\;\;
\Phi_2
=\frac{e^{i\rho}}{c\sqrt{2}}
\left(\frac{\lambda_1+\lambda_3}{\lambda_2+\lambda_3}\right)^\frac{1}{4}
\left(\begin{array}{c}0\\v_1\end{array}\right)
\end{equation}
with $c$ real and get
\begin{equation}
V=\begin{array}{c}\left[\left(1+
\frac{\lambda_3}{\sqrt{(\lambda_1+\lambda_3)(\lambda_2+\lambda_3)}}
\right)\frac{(c-1)^4}{c^2}
+\left(1-\frac{\lambda_3}{\sqrt{(\lambda_1+\lambda_3)(\lambda_2+\lambda_3)}}
\right)\frac{(c+1)^2(c-1)^2}{c^2}\right]
\frac{\lambda_1+\lambda_3}{8}v_1^4\end{array}.
\end{equation}
For $c\neq1$ sufficiently close to 1, the second term dominates and again
we need (\ref{dbl3nec1}).

If $\lambda_1+\lambda_3=0$ and/or $\lambda_2+\lambda_3=0$, take
\begin{equation}
\label{dbphil123c}
\Phi_1
=\frac{c}{\sqrt{2}}
\left(\begin{array}{c}0\\v_1\end{array}\right),
\;\;\;\;\;\;
\Phi_2
=\frac{e^{i\rho}}{c\sqrt{2}}
\left(\begin{array}{c}0\\v_2\end{array}\right)
\end{equation}
with $c$ real and get
\begin{equation}
\label{dbv123d}
V=(\lambda_1+\lambda_3)(c^2-1)^2\frac{v_1^4}{4}
+(\lambda_2+\lambda_3)\left(1-\frac{1}{c^2}\right)^2\frac{v_2^4}{4}
-2\lambda_3(c^2-1)\left(1-\frac{1}{c^2}\right)\frac{v_1^2v_2^2}{4}.
\end{equation}
If $\lambda_1+\lambda_3>0$ and $\lambda_2+\lambda_3=0$
($\lambda_1+\lambda_3=0$ and $\lambda_2+\lambda_3>0$), the last term in
(\ref{dbv123d}) is dominant for small enough (large enough) $c^2$ and
$\lambda_3\leq0$, i.e.\ again (\ref{dbl3nec1}), is necessary. For
$\lambda_1+\lambda_3=\lambda_2+\lambda_3=0$, any $c^2\neq0,1$ leads to this
conclusion. Therefore (\ref{dbl3nec1}) is necessary in any case.

Vacuum stability requires $V$ to be nonnegative for large fields,
where di\-men\-sion-two terms can be neglected.
Keeping only quartic terms and writing \mbox{$a\equiv\Phi_1^\dagger\Phi_1$},
$b\equiv\Phi_2^\dagger\Phi_2$, $x\equiv{\rm Re}(\Phi_1^\dagger\Phi_2)$,
and $y\equiv{\rm Im}(\Phi_1^\dagger\Phi_2)$, where the restrictions
\mbox{$a$, $b\geq0$} and $x^2+y^2\leq ab$ apply, we get
\begin{equation}
V=(\lambda_1+\lambda_3)a^2+(\lambda_2+\lambda_3)b^2
+(2\lambda_3+\lambda_4)ab+(\lambda_5-\lambda_4)x^2
+(\lambda_6-\lambda_4)y^2.
\end{equation}
For given $ab$, the sum of the last two terms is easily seen to have a
minimum value of $(\lambda_<-\lambda_4)ab$, where $\lambda_<$ is defined as
\begin{equation}
\lambda_<\equiv\min(\lambda_4,\lambda_5,\lambda_6).
\end{equation}
We therefore have to minimize
\begin{equation}
\label{dbvab}
V=\left(\sqrt{\lambda_1+\lambda_3}\,a
-\sqrt{\lambda_2+\lambda_3}\,b\right)^2
+\left(2\lambda_3+\lambda_<
+2\sqrt{(\lambda_1+\lambda_3)(\lambda_2+\lambda_3)}\,\right)ab
\end{equation}
in order to see if $V\geq0$ for all large fields $\Phi_1$, $\Phi_2$.
It is easy to convince oneself that even if $\lambda_1+\lambda_3=0$ and/or
$\lambda_2+\lambda_3=0$, (\ref{dbvab}) leads to the necessary condition
\begin{equation}
\label{dbl3nec2}
\lambda_3\geq-\left(\lambda_</2
+\sqrt{(\lambda_1+\lambda_3)(\lambda_2+\lambda_3)}\,\right)
\end{equation}
to prevent $V$ from becoming negative for some large fields.

So far we have derived the necessary conditions (\ref{dbl4nec}),
(\ref{dbl5nec}), (\ref{dbl6nec}), (\ref{dbl123nec}),
(\ref{dbl3nec1}), and (\ref{dbl3nec2}). Now we will show that the same
conditions are sufficient as well. For this purpose, we will assume from
now on that the necessary conditions are fulfilled. Note that one of the
consequences is that $V_{456}$, i.e.\ the part of the potential involving
$\lambda_{4,5,6}$, is never negative. The same is not necessarily true for
the other part $V_{123}$ of the potential.

If both $\lambda_1+\lambda_3>0$ and $\lambda_2+\lambda_3>0$, we can write
\begin{eqnarray}
V&\!\!\!\!=\;&\!\!\!\begin{array}{l}\frac{1}{2}\left(
1+\frac{\lambda_3}{\sqrt{(\lambda_1+\lambda_3)(\lambda_2+\lambda_3)}}
\right)\!\left[
\sqrt{\lambda_1+\lambda_3}\left(\Phi_1^\dagger\Phi_1-\frac{v_1^2}{2}\right)
+\sqrt{\lambda_2+\lambda_3}\left(\Phi_2^\dagger\Phi_2-\frac{v_2^2}{2}\right)
\right]^2\end{array}
\nonumber\\
&+\!\!&\!\!\!\begin{array}{l}\frac{1}{2}\left(
1-\frac{\lambda_3}{\sqrt{(\lambda_1+\lambda_3)(\lambda_2+\lambda_3)}}
\right)\!\left[
\sqrt{\lambda_1+\lambda_3}\left(\Phi_1^\dagger\Phi_1-\frac{v_1^2}{2}\right)
-\sqrt{\lambda_2+\lambda_3}\left(\Phi_2^\dagger\Phi_2-\frac{v_2^2}{2}\right)
\right]^2\end{array}\nonumber\\
&+\!\!&\!V_{456}.
\end{eqnarray}
Thus
\begin{equation}
\label{dbl3suff1}
|\lambda_3|\leq\sqrt{(\lambda_1+\lambda_3)(\lambda_2+\lambda_3)}
\end{equation}
is sufficient to have $V_{123}\geq0$. If $\lambda_1+\lambda_3=0$ and/or
$\lambda_2+\lambda_3=0$, then manifestly $V_{123}\geq0$ if (\ref{dbl3suff1})
holds, i.e.\ if $\lambda_3=0$. Since $V_{456}\geq0$, (\ref{dbl3suff1})
is sufficient to have $V\geq0$.

Define $x_1\equiv\sqrt{\Phi_1^\dagger\Phi_1}+v_1/\sqrt{2}$,
$y_1\equiv\sqrt{\Phi_1^\dagger\Phi_1}-v_1/\sqrt{2}$,
$x_2\equiv\sqrt{\Phi_2^\dagger\Phi_2}+v_2/\sqrt{2}$, and
$y_2\equiv\sqrt{\Phi_2^\dagger\Phi_2}-v_2/\sqrt{2}$.
Making use of the necessary conditions, it follows that
\begin{eqnarray}
V&\geq&V_{123}+\lambda_<
\left\{(\Phi_1^\dagger\Phi_1)(\Phi_2^\dagger\Phi_2)
-(\Phi_1^\dagger\Phi_2)(\Phi_2^\dagger\Phi_1)
\invexpa\right.\nonumber\\
&&\hspace{60pt}\left.
+\left[{\rm Re}(\Phi_1^\dagger\Phi_2)-\frac{v_1v_2}{2}\cos\rho\right]^2
+\left[{\rm Im}(\Phi_1^\dagger\Phi_2)-\frac{v_1v_2}{2}\sin\rho\right]^2
\right\}\nonumber\\
&=&V_{123}+\lambda_<\left[(\Phi_1^\dagger\Phi_1)(\Phi_2^\dagger\Phi_2)
-2{\rm Re}\left(\Phi_1^\dagger\Phi_2e^{-i\rho}\frac{v_1v_2}{2}\right)
+\frac{v_1^2v_2^2}{4}\right]\nonumber\\
&\geq&V_{123}
+\lambda_<\left[\sqrt{(\Phi_1^\dagger\Phi_1)(\Phi_2^\dagger\Phi_2)}
-\frac{v_1v_2}{2}\right]^2\nonumber\\
&=&(\lambda_1+\lambda_3)x_1^2y_1^2+(\lambda_2+\lambda_3)x_2^2y_2^2
+2\lambda_3x_1y_1x_2y_2+\frac{\lambda_<}{4}[x_1y_2+x_2y_1]^2\nonumber\\
&=&\left(\sqrt{\lambda_1+\lambda_3}\,x_1y_1
-\sqrt{\lambda_2+\lambda_3}\,x_2y_2\right)^2\nonumber\\
&&-\frac{1}{2}\left[
\lambda_3+\sqrt{(\lambda_1+\lambda_3)(\lambda_2+\lambda_3)}\right]
(x_1y_2-x_2y_1)^2\nonumber\\
&&+\frac{1}{2}\left[\frac{\lambda_<}{2}
+\lambda_3+\sqrt{(\lambda_1+\lambda_3)(\lambda_2+\lambda_3)}\right]
(x_1y_2+x_2y_1)^2.
\end{eqnarray}
Thus, obviously
\begin{equation}
\label{dbl3suff2}
-\left(\lambda_</2+\sqrt{(\lambda_1+\lambda_3)(\lambda_2+\lambda_3)}\,\right)
\leq\lambda_3\leq-\sqrt{(\lambda_1+\lambda_3)(\lambda_2+\lambda_3)}
\end{equation}
is sufficient to have $V\geq0$.

Combining (\ref{dbl3suff1}) and (\ref{dbl3suff2}), we get the new sufficient
condition
\begin{equation}
-\left(\lambda_</2+\sqrt{(\lambda_1+\lambda_3)(\lambda_2+\lambda_3)}\,\right)
\leq\lambda_3\leq\sqrt{(\lambda_1+\lambda_3)(\lambda_2+\lambda_3)}.
\end{equation}
However from (\ref{dbl3nec1}) and (\ref{dbl3nec2}) we already know that
this is also necessary.

Let us summarize the bounds on the parameters in the potential that
are both necessary and sufficient to ensure that $V\geq0$ for all fields
$\Phi_1$ and $\Phi_2$, thereby ensuring vacuum stability and---up to possible
degeneracy of different vacua---the desired symmetry breaking pattern. They
are the remarkably simple conditions
\begin{equation}
\label{db13}
\lambda_1+\lambda_3\geq0,\;\;\;\;\lambda_2+\lambda_3\geq0,\;\;\;\;
\lambda_3\geq-\left[\frac{1}{2}\min(\lambda_4,\lambda_5,\lambda_6)
+\sqrt{(\lambda_1+\lambda_3)(\lambda_2+\lambda_3)}\right],
\end{equation}
\begin{equation}
\label{db3b}
\lambda_3\leq\sqrt{(\lambda_1+\lambda_3)(\lambda_2+\lambda_3)},
\;\;\;\;\lambda_4\geq0,\;\;\;\;\lambda_5\geq0,\;\;\;\;\lambda_6\geq0.
\end{equation}
Using (\ref{db567phi}) and (\ref{db56phi}), these conditions can easily be
translated to the case of nonzero $\lambda_7$.

Now up to marginal cases, the quartic terms in the potential alone decide
about vacuum stability. Then the first three conditions
(\ref{db13}) follow from vacuum stability requirements, while
the last four bounds (\ref{db3b}) ensure the correct symmetry
breaking pattern by the vacuum. This distinction is interesting because
if the couplings are made running through the renormalization group,
one does not want to loose vacuum stability of the two-doublet
potential at scales before new physics comes in, while the other
conditions are important only for small fields, and therefore their
behavior at large scales through running is irrelevant.

\section*{Acknowledgements}
This work was supported in part by the Department of Energy under Contract
No.~DE-AT03-88ER 40383 Mod A006-Task C.

\end{document}